\documentclass{aa}
\usepackage{graphics}

\def\ltapprox{\raise 2pt \hbox {$<$} \kern-1.1em \lower 5pt \hbox {$\approx$}}
\def\ltsim{\raise 2pt \hbox {$<$} \kern-1.1em \lower 4pt \hbox {$\sim$}}
\def\gtsim{\raise 2pt \hbox {$>$} \kern-1.1em \lower 4pt \hbox {$\sim$}}

\newcommand{\apj}[2]{\mbox{{\em Ap.J.,\ }{ #1},{ #2}}}

\newcommand{\aaa}[2]{\mbox{ {\em A.\&A.,\ }{ #1},{ #2}}}

\newcommand{\mnras}[2]{\mbox{ {\em M.N.R.A.S., }{ #1},{ #2}}}

\newcommand{\apjl}[2]{\mbox{ {\em Ap.J.Letters,\ }{ #1},{ #2}}}
\newcommand{\apss}[2]{\mbox{ {\em Astroph. \&Sp. Science,\ }{ #1},{ #2}}}
\newcommand{\jgr}[2]{\mbox{ {\em Journ. Geophys. Res.,\ }{ #1},{ #2}}}

\begin{document}

\titlerunning{Astrophysical UI powered by GW emission
}
\title{Astrophysical Unipolar Inductors powered by GW emission}

\author{S.Dall'Osso\inst{1}
\and G.L.Israel\inst{1}
\and L.Stella\inst{1} 
}

\offprints{dallosso@mporzio.astro.it}

\institute{
INAF--Osservatorio Astronomico di Roma, via Frascati 33, I--00040 Monteporzio 
Catone
(Roma), Italy; dallosso, gianluca and stella@mporzio.astro.it 
%\and
%Affiliated to the International Center for Relativistic Astrophysics
}
\date{}

\abstract{
We consider the Unipolar Inductor Model (Goldreich \& Lynden-Bell 
1969) applied to Double Degenerate Binaries (DDBs) with ultrashort periods 
(Wu et al. 2002). \\
In this model a magnetized primary white dwarf has a slight asynchronism 
between its spin and orbital motion, so that the (non-magnetic) secondary 
experiences a motional electric field when moving through the primary field 
lines. This induces a current flow between the two stars and provides
an electric spin-orbit coupling mechanism for the primary.\\
We study the combined effect of Gravitational Wave emission and
electric spin-orbit coupling on the evolution of the primary degree of 
asynchronism and the associated rate of electric current dissipation in such 
systems, assuming that the primary's spin is not affected 
by other mechanisms such as tidal interactions with the companion. In 
particular, we show that in ultrashort period binaries the emission of GW pumps
 energy in the electric circuit as to keep it steadily active.
This happens despite the fact that spin-orbit coupling can rapidly synchronize 
the primary, because GW represent a slow desynchronizing mechanism that
steadily substracts orbital angular momentum to the system. A slightly 
asynchronous steady-state is thus achieved, determined by the balance between 
these two competing effects. This can be shown to correspond to a condition
where the total available electric energy is conserved, because of GW emission,
 while dissipation, synchronization and orbital shrinking continue.
}
\maketitle

\section{Introduction}
\label{intro}
The Unipolar Inductor Model (UIM from here on) has been originally proposed
(Goldreich \& Lynden-Bell 1969) for the Jupiter-Io system to explain the 
origin of bursts of decametric radiation received from the planet, whose 
occurrence and intensity were known to be strongly influenced by the 
orbital location of Io.\\ 
The model relies on the fact that Jupiter's spin differs from the system 
orbital period, Io's spin being locked to the orbit around the planet. Given 
the $\sim$ 10 G surface dipole magnetic field of Jupiter and the good 
electrical conductivity expected from the satellite, the system behaves as a 
remarkably simple electric circuit, that can be sketched as follows. A good 
conductor (the satellite) has a trasverse motion with respect to the field 
lines of the planet and this induces an e.m.f. across it\footnote{It would be 
more appropriate to describe this in terms of the Lorenz force acting on the 
charge carriers within the conductor, but introducing the e.m.f. emphasises the
 analogy with an electrical circuit}. The e.m.f. can accelerate free charges 
that are present in the ambient medium, giving rise to a flow of currents 
between the two objects. Currents are confined to a thin sheet along the 
sides of the flux tube connecting Jupiter and Io; hence, the cross section of 
the current layers at Jupiter's surface has the form of two arc-shaped strips. 
Currents flow along one side of the tube towards Jupiter and then vertically 
through its upper atmosphere and ionosphere.
Jupiter's electrical conductivity ($\sigma$) in the ionosphere is expected to 
be almost isotropical (Goldreich \& Lynden-Bell 1969 and references therein): 
hence, currents reaching it can propagate transverse to field lines, 
closing the circuit and returning back to Io along the opposite side of the 
flux tube. Charges along the flux tube are accelerated to mildly relativistic 
energies and lead to coherent cyclotron emission over a range of wavelenghts: 
this is the framework in which Jupiter's decametric radiation and its strong 
modulation by Io's position can be explained.\\
Among the main expectations of the model it is of interest here mentioning the
effect of resistive dissipation of currents in the planetary atmosphere: this
causes a local heating and an associated localized enhancement of the thermal 
emission.\\
Several confirmations to the model have been obtained over the years, the most 
spectacular being provided by the HST UV observations of Clarke et al. (1996). 
In particular, these revealed the emission from the resistive dissipation 
region on Jupiter's surface - Io's footprint - and its corotation with the 
satellite, which means that the emission region drifts on the planet's surface 
exactly as expected in the UIM.\\
The same basic picture has been proposed by Li et al. (1998) for planetary 
companions to white dwarfs: if the UIM applied to such systems it would offer 
an alternative way of searching for extrasolar planets through the 
electromagnetic emission associated to the electrical circuit.\\
Wu et al. (2002) proposed a similar scenario for the case of two white dwarfs 
forming a close binary system. They consider a moderately magnetized primary, 
a non-magnetic secondary and a primary spin not perfectly synchronous with the 
orbital motion, while the secondary's is efficiently kept synchronous by tidal 
forces. The atmospheric electrical conductivity of white dwarfs is expected
to be highly anisotropic, because of the WD magnetic field, up to at least 
$\sim$ 1 Km depth (Li, Ferrario and Wickramashige 1998), so currents are 
forced to follow field lines in this situation: they can only cross them and 
close the circuit when $\sigma$ becomes isotropical and cross-field diffusion 
can proceed efficiently.\\ 
This model was proposed to account for the observed properties of the two 
candidate ultrashort period binaries RX J0806+15 and RX J1914+24. These have
significant soft X-ray emission ($\sim 10^{32}~\mbox{and}~ 10^{34}$ erg 
s$^{-1}$ respectivley) pulsed at periods $\sim$ 321 and 569 s: these 
periodicities are interpreted as likely due to orbital modulations, although 
the subject is still debated (see Cropper et al. 2003 for a review on this and 
several other aspects). Further, both sources have a measured orbital spin-up 
(Strohmayer 2004, Hakala et al. 2004, Israel et al. 2004), as if driven by GW 
with no matter transfer: this would thus require an alternative 
source for the X-ray emission, UIM representing an interesting 
possibility. In fact, given the very short orbital periods, possibly larger 
primary magnetic fields and the compactness of both components with respect to 
previous versions of the UIM, such systems may dissipate energy at a 
significant rate even for slight degrees of asynchronism, as shown by Wu et 
al. (2002).
These authors showed that currents are resistively dissipated essentially in 
the primary atmosphere, the associated heating causing the observed soft X-ray 
emission: its source is ultimately represented by the energy of the relative 
motion between the primary spin and the orbit. Further, the process 
redistributes angular momentum - via the Lorentz torque on cross-field 
currents - between the primary spin and the orbit and thus acts to 
synchronize them.\\
In this application of the UIM, synchronization timescales $\sim$ few 10$^3$ 
yrs were obtained for both RX J1914+24 and RX J0806+15, very short compared to 
the orbital evolutionary timescales, respectively $\sim 8 \times 10^6$ yrs and 
3$\times 10^5$ yrs. Accordingly, a very low probability (significantly $<$ 
1\%) of detecting these systems during the asynchronous phase would be 
expected. This would in turn require a very large population of such systems 
in the Galaxy, much larger than predicted by population-synthesis models, 
since two of them have been detected in the short-lived, active phase. \\
Concerning this point, we focus in this work on a key aspect of the problem 
that has been overlooked in previous works, namely that in the framework of 
the UIM applied to DDBs perfect synchronization between the primary spin and 
the orbit is \textit{never} reached: therefore, the current flow should not 
stop at any time. This happens even if the synchronization timescale 
($\tau_{\alpha}$) is much shorter than the orbit evolutionary timescale 
($\tau_o = \omega_o / \dot{\omega}_o$), because of the continuous loss of 
orbital angular momentum caused by GW. As tidal synchronization is expected to 
be efficient only for the 
lighter star, GW drive the primary out of synchronism on the longer timescale 
$\tau_o$, thus continuosly feeding energy to the electric circuit.\\
Consider a perfectly synchronous system: the electric circuit would be turned 
off while GW would still cause orbital spin-up and, thus, a desynchronization 
of the primary's spin, which would in turn switch the circuit on: the system 
should thus evolve, over the timescale $\tau_{\alpha}$, towards a slightly 
asynchronous steady-state that is determined by the balance between the fast, 
synchronizing mechanism (UI) and the slow, desynchronizing one (GW).\\
In the following this issue will be addressed quantitatively: in particular
we show that, in a binary system with UIM at work, the orbital period 
decreases because of GW emission and the primary spin is forced by the 
coupling to approach the orbital period, but perfect equality is not reached 
because of the energy fed by GWs to the circuit. Further, we derive a 
definition of $\tau_{\alpha}$ as a function of system parameters, find 
conditions under which $\tau_{\alpha} \ll \tau_o$ is verified and discuss the
salient evolutionary features of the UIM implied by this condition.\\
In a companion paper we apply these general considerations - and the related
formulas that we derive here - to RX J0806+15 and RX J1914+24 and obtain 
interesting constraints on system parameters for the UIM to be applicable to 
these sources. \\
Detailed and systematic calculations and evolutionary implications are beyond 
the scope of these works and will be addressed elsewhere.
\section{Asynchronous Evolution in the Unipolar Inductor Model}
\label{general}
Following Wu et al. (2002), the primary asynchronism parameter is defined as
$\alpha = \omega_1/ \omega_o$, where $\omega_1$ is the primary spin and 
$\omega_o$ the orbital motion.\\
Given an asynchronous system with orbital separation $a$, the secondary star
will be moving across the primary magnetic field lines with the relative 
velocity $ v = a 
(\omega_o - \omega_1) = [G M_1 (1+q)]^{1/3}~\omega^{1/3}_o (1-\alpha)$,
 where $G$ is the gravitational constant, $M_1$ the primary mass,
$q = M_2/M_1$ the system mass ratio. The electric field induced through the 
secondary is {\boldmath$E$} = $\frac{\mbox{{\boldmath$v \times B$}}}{c}$, with 
an associated e.m.f. $\Phi = 2R_2 E$, $R_2$ being the secondary's radius. 
Because of the induced e.m.f. a current flows between the two component stars, 
whose resistive dissipation - which takes place essentially in the primary 
atmosphere - has two effects: first, it causes significant heating of the 
dissipation region, thus powering its soft X-ray emission. 
Second, the Lorentz torque associated to currents crossing field lines in the 
primary atmosphere (and in the conducting secondary) redistributes angular 
momentum between the primary spin and the orbital motion. In presence of 
significant GW-emission, as expected for two white dwarfs orbiting each other 
at ultrashort periods, there is an additional effect: orbital 
angular momentum is continuously lost from the system, causing the orbital 
period to steadily decrease (and the orbit to shrink). 
Hence, as long as the primary is not efficiently kept synchronous by tidal 
forces, its spin will lag behind the orbital motion on the orbital 
evolutionary timescale; electric coupling will thus be active and exchanging 
angular momentum between the orbit and the primary spin.\\
We stress that the absence of any mechanism other than the Unipolar
Inductor able to affect the primary's spin over the orbit evolutionary 
timescale is an essential assumption of our analysis: in Appendix A we show
 that tidal synchronization of the primary is indeed not expected to be 
efficient on 
this timescale, while it may well be effective in rapidly synchronizing a 
low-mass companion. The latter effect is discussed in Appendix A as
well, being however of much lower relevance.\\
In this work we consider binary systems consisting of two degenerate white 
dwarfs, with the following mass-radius relation (Nauenberg 1972, Marsh 
\& Steeghs 2002):
\begin{equation}
\label{massradius}
\frac{R}{R_{\odot}} = 0.0112 \left[ \left(\frac{M}{1.433}\right)^{-\frac{2}{3}}
- \left(\frac{M}{1.433}\right)^{\frac{2}{3}}\right]^{\frac{1}{2}}
\end{equation}
where $M$ is expressed in solar masses.\\
In order to make all above features more quantitative we need proper
expressions for the physical quantities of interest. Let us start 
recalling eq. E4 from Appendix E of Wu et al. (2002), that describes the 
evolution of the orbital period of a binary system with the coupled effects of 
UIM and GW:
\begin{equation}
\label{omegadot}
\frac{\dot{\omega}_o}{\omega_o} = \frac{1}{g(\omega_o)}\left(\dot{E}_g
- \frac{W}{1-\alpha}\right)
\end{equation}
Here $\dot{E}_g = - (32/5)G/c^5 [q/(1+q)]^2 M^2_1 a^4 \omega^6_o$ is 
the energy loss rate through GW emission (Landau \& Lifshitz 1951), the second 
term in parenthesis represents the contribution of electric coupling between 
the primary spin and the orbit and W is the electric current dissipation 
rate (source luminosity). The function $g(\omega_o) < 0$ is:
\begin{equation}
\label{gomega}
g(\omega_o) = -\frac{1}{3} \left[ \frac{q^3 G^2 M^5_1 \omega^2_o}{(1+q)} 
\right]^{\frac{1}{3}} \left[1-\frac{6}{5} (1+q) \left(\frac{R_2}{a}\right)^2
\right]
\end{equation}
The first term (including the coefficient -1/3) corresponds to the ratio 
between $\dot{E}_g$ and $\dot{\omega}_o/ \omega_o$ for two point masses with 
no spin-orbit coupling of any kind. It represents two thirds of the total 
orbital (gravitational plus kinetic) energy of the binary system ($E_g$). The 
second term in square brackets - call it $h(a)$ - is of 
order unity for most plausible system parameters: in particular, 0.6 $ < h(a) 
< 1$ for orbital periods longer than $200$ s, primary mass M$_1 > 0.4$ 
M$_{\odot}$ and secondary mass M$_2 > 0.08$ M$_{\odot}$ but, unless extreme 
values of all parameters at the same time are assumed, it is $> 0.85$ in most 
cases. 
This term accounts for the 
secondary spin being tidally locked to the orbit; as the system shrinks due to 
GW emission, an additional tiny amount of orbital angular momentum is lost to 
spin up the secondary and the resulting $\dot{\omega}_o / \omega_o$ will be 
just slightly higher, given $M_1, q$ and $\omega_o$. Eq. 
(\ref{gomega}) can thus be written concisely as $g(\omega_o) = (2/3) E_g h(a)$,
a physically much clearer expression which will be frequently used especially 
in $\S$ \ref{energy}. 
As both $\dot{E}_g$ and $g(\omega_o)$ are negative, GW always give a 
positive contribution to $\dot{\omega}_o$ while electric coupling ($W > 0$ by 
definition) may either favour or oppose the orbital spin-up depending on the 
sign of $(1-\alpha)$.\\
Since the two candidate ultrashort period binaries have a measured orbital
spin-up, in the following we will be interested in systems where the orbit
is shrinking: if $\alpha<1$ this is warranted, but when $\alpha>1$ spin-orbit 
coupling transfers angular momentum \textit{to} the orbit and, if sufficiently
strong, it may even overcome the effect of GW. \\
From a general point of view, however, there is no a priori reason not to 
consider systems subject to such a phase of orbital spin-down, due to a strong 
spin-orbit coupling.
In $\S$\ref{conserving} and Appendix B we briefly comment on this 
situation. For the moment, however, we focus only on systems where 
$\dot{\omega}_o > 0$, a condition that, if $\alpha >1$, must be implicitly 
expressed as $|\dot{E}_g| > |W/(1-\alpha)|$ and will be made more explicit in 
the next sections.
\\
The quantity W can be expressed as: 
%(see Wu et al. 2002):
%
\begin{equation}
\label{W}
W  =  \frac{\Phi^2}{\Re} = \left(\frac{\mu_1 R_2}{c}\right)^2 
\frac{[G M_1 (1+q)]^{-\frac{4}{3}}}{\Re}~\omega^{14/3}_o 
(1-\alpha)^2
\end{equation}
where $\mu_1$ is the primary magnetic moment and the system's effective 
resistance $\Re$ is (see Wu et al. 2002):
\begin{equation}
\label{resistance}
\Re = \frac{1}{2\overline{\sigma} R_2} \left(\frac{H}{\Delta d}\right)\jmath(e)
\left(\frac{a}{R_1}\right)^{3/2} = N \omega^{-1}_o
\end{equation}
where $N$ includes $G$, $M_1$ and $q$ after writing $a$ according to Kepler's 
third law.
In the above formula $\overline{\sigma}$ is the height averaged WD atmospheric 
conductivity, $H$ the atmospheric depth at which currents cross magnetic field 
lines and return back to the secondary and $\Delta d$ the thickness of the 
arc-like cross section of the current layer at the primary atmosphere. 
Finally, the geometric factor $\jmath(e) \sim 1$ when the orbital period 
is less than 1 hr or so (see Wu et al. 2002). 
Combining the two above relations we obtain:
\begin{eqnarray}
\label{Wrewritten}
W & = & \left(\frac{\mu_1}{c}\right)^2 \frac{2 \overline{\sigma} R^{3/2}_1 
R^3_2}{[GM_1(1+q)]^{11/6}} \frac{\omega^{17/3}_o (1-\alpha)^2}{(H /\Delta d) 
\jmath(e)} \nonumber \\
& = & k~\omega^{\frac{17}{3}}_o (1-\alpha)^2
\end{eqnarray}
where the last equality defines $k$.\\
Finally, combining eq. E4 and E5 from Appendix E of Wu et al. (2002), the 
following expression for the evolution of $\alpha$ is obtained:
\begin{eqnarray}
\label{alfaevolution}
\frac{\dot{\alpha}}{\alpha} & = & \frac{\dot{\omega}_1}{\omega_1} - 
\frac{\dot{\omega}_o}{\omega_o} = - \frac{\dot{\omega}_o}{\omega_o} +
\frac{W}{\alpha(1-\alpha) I_1 \omega^2_o} \Rightarrow \nonumber \\
\dot{\alpha} & + & \left(\frac{\dot{\omega}_o}{\omega_o} + \frac{k}{I_1} 
\omega^{\frac{11}{3}}_o \right) \alpha - \frac{k}{I_1} 
\omega^{\frac{11}{3}}_o = 0
\end{eqnarray}
where $I_1$ is the moment of inertia of the primary star.
\subsection{General Considerations and Approximations}
\label{consider}
Eq. (\ref{alfaevolution}) in its general form is a non linear, first order 
differential equation for $\alpha$, with time-dependent coefficients, the 
non-linear term being due to the coupling between UI and GW in the orbital 
evolution, \textit{i.e.} by the torque that exchanges angular momentum between 
the primary spin and the orbit. \\
Independent of the exact solution, a general remark can be made 
that provides a better understanding of the problem and allows finding 
approximations that simplify its mathematical treatment without 
affecting its very nature. \\
The key point is that the orbital period changes continuously over time: hence,
 synchronization would be maintained only if the primary spin changed 
continuously as well. If UI was the only mechanism affecting the primary spin
- as we are assuming - the electrical circuit would have to remain 
always active in order to allow $\omega_1$ to track $\omega_o$.\\
Note that this conclusion holds in general, whatever the functional form of 
$\dot{\omega}_o$ is. On the other hand, the 
details of how the system evolves and the state to which it is led depend on 
the functional form of $\dot{\omega}_o$, \textit{i.e.} on the model adopted.\\
In the following we obtain an analytic solution to the evolutionary equation
in a specific approximation of the UIM and show that it implies the existence 
of a slightly asynchronous, asymptotic state\footnote{The meaning of this will 
be extensively discussed in the rest of the paper.}. 
The physical conditions under which our particular solution applies are 
described in detail, but we stress that the conclusion concerning 
the existence of a slightly asynchronous, asymptotic state should have a more 
general validity.\\
The most natural choice to begin with is to focus on the 
simplified problem in which the effects of UI and GW on the evolution of the 
asynchronism can be decoupled, thus neglecting the non-linear term
in the evolutionary equation. The primary spin evolution is 
clearly independent of GW and is driven by UI alone; so let
 us specialize to the case where GW alone drive the orbital evolution, UI 
providing only a negligible contribution to it. 
\subsubsection{Validity of the Approximation}
\label{check}
\begin{figure}[h]
\includegraphics{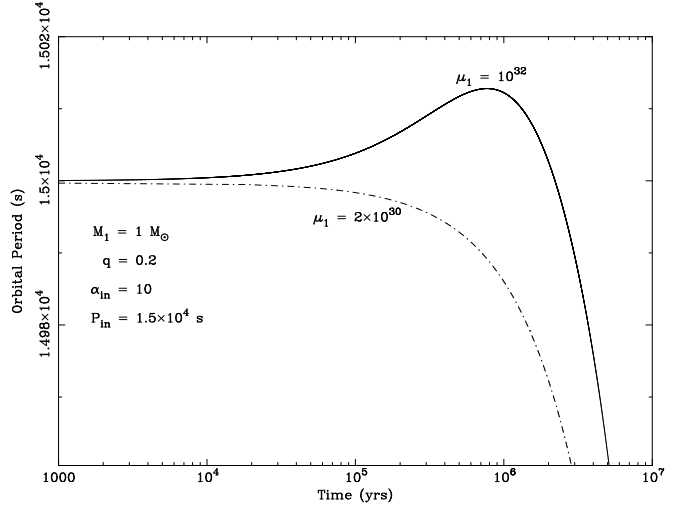}
\vspace{6.6cm}
\caption{Illustrative example of the early orbital evolution of a DDB subject 
to the UI mechanism. The plots are obtained through a numerical integration of 
the coupled evolutionary equations for $\alpha$ and $\omega_o$. Initial 
conditions are chosen quite arbitrarily, fo illustrative purposes only: the 
initial orbital period is P$_{in} = 1.5 \times 10^4$ s ($\sim$ 4 hrs) and the 
primary initial spin $\sim$ 25 min ($\alpha_{in} = 10$). Other system 
parameters are indicated in the figure. Two primary magnetic moments were 
tried and their values, in G cm$^3$, label the corresponding curves.}
\label{period}
\end{figure}

We check the relative magnitudes of the terms\footnote{more 
precisely, we compare their absolute values} in eq. (\ref{alfaevolution}) to 
determine the conditions that allow this approximation to be 
introduced consistently.
%the neglect of the non-linear term, \textit{i.e.} to 
%decouple UI and GW. 
%Re-write eq. 
Using eq. (\ref{omegadot}) to re-express $\dot{\omega}_o / \omega_o$ in 
(\ref{alfaevolution}) it obtains:
\begin{equation}
\label{alfaevolutionextended}
\frac{\dot{\alpha}}{\alpha} = - \underbrace{\frac{\dot{E}_g}{g(\omega_o)}}_{A} 
+ \underbrace{\frac{W}{g(\omega_o) (1-\alpha)}}_{B} + 
\underbrace{\frac{W}{\alpha (1-\alpha) I_1 \omega^2_o}}_{C}
\end{equation}
Let us first compare (B), the contribution of UI to the orbital evolution to 
(C), the contribution of UI to the primary spin evolution.
The condition $B \ll C$ (say, $B < 10^{-1} C$) will be met if:
\begin{equation}
\label{BlessthanC}
%\frac{W}{g(\omega_o) (1-\alpha)} & < & \frac{1}{10} \frac{W}{\alpha 
%(1-\alpha) I_1 \omega^2_o}~~\Longleftrightarrow~~\frac{I_1 \omega^2_o}
%{g(\omega_o)} <
%\frac{1}{10 \alpha} \nonumber \\
\alpha < \frac{g(\omega_o)}{10~I_1 \omega^2_o} \propto \omega^{-\frac{4}{3}}_o 
\end{equation}
where the right-hand term is a decreasing function of $\omega_o$ and, for 
component masses in the range ($M_1 \sim 0.5 \div 1$ M$_{\odot}$), gives 
$\alpha <2$ even at an orbital period $P_o \sim$ 300 s. Therefore, if $P_{orb} 
> 300$ s the condition $B \ll C$ is met for most plausible values of the 
asynchronism.\\
We still need to determine the conditions under which
%prove that the condition 
$ B \ll A$ is true as well and, then, that GW mostly determine the 
orbital evolution. By requiring $B < 10^{-1} A$ we obtain:
\begin{eqnarray}
\label{condition}
\omega_o  <  \omega_{GW} & = & 8.5 \times 10^{-27} q^{6/7} (1+q)^{1/2}
\frac{M^{31/14}_1}{(2 \sigma R^3_2 R^{3/2}_1)^{3/7}} \nonumber \\
 & & \times \frac{(H/ \Delta d)^{3/7}}{(1-\alpha)^{3/7}~\mu^{6/7}_1}~~~
\mbox{ or} \nonumber \\
P_o > P_{GW} & = & 410 \frac{1}{q^{\frac{6}{7}} (1+q)^{\frac{1}{2}}}
\frac{\left(\frac{R_2}{1.7 \times 10^9}\right)^{\frac{9}{7}} 
\left(\frac{R_1}{6 \times 10^8}\right)^{\frac{9}{14}}} {\left(\frac{M_1}
{0.9~M_{\odot}}\right)^{\frac{31}{14}}} \nonumber \\
 & & \times \left(\frac{\mu_1}{2 \times 10^{30}}\right)^{\frac{6}{7}}
 \left[\frac{H}{\Delta d} \jmath(e) \right]^{-\frac{3}{7}}(1-\alpha)
^{\frac{3}{7}}
\end{eqnarray} 
where c.g.s. units are omitted in the normalizations and a height-averaged 
conductivity $\overline{\sigma} = 3\times 10^{13}$ e.s.u. has been assumed
(Wu et al. 2002).
Here $\mu_1$ has been normalized to a low value because this gives
more easily shrinking orbits. Indeed, it can be directly checked from the above
equation that, if $\mu_1 \sim 10^{32}$ G cm$^3$
as suggested originally by Wu et al. (2002), $P_{GW} \geq 10^4$ s even for 
$(1-\alpha) \sim 0.1$. This can be stated as follows: relatively highly
magnetized systems with $\alpha >1$ do not shrink unless their 
degree of asynchonism is very low, because spin-orbit 
coupling dominates over GW emission. They can shrink only if either they are 
formed with $\alpha$ very close to one (which would require some spin-orbit 
coupling even during the common envelope phase that likely leads to their 
formation) or begin their evolution with $\alpha <1$, quite a remarkable slow 
spin if one considers an initial orbital period $\sim 10^4$ s. What is more 
likely is that at an orbital period of, say, 5 hr the primary spin is 
faster than that (say, $\alpha \sim$ a few). 
In this case, a $10^{32}$ G cm$^3$ magnetic moment would cause the orbit to 
widen until $(1 - \alpha)\omega^{17/3}_o$ is small enough that GW 
dominate over spin-orbit coupling in $\dot{\omega}_o$.\\ 
Let us restrict attention to weakly magnetized systems ($\mu_1 < 10^{31}$
G cm$^3$): even in this case, condition (\ref{condition}) is more constraining 
than (\ref{BlessthanC}): it can be met by low mass binaries, for 
$\mu_1 \sim$ a few $10^{30}$ G cm$^3$ and P $\leq$ 2000 s, only if $(1-\alpha)$
 is quite small ($<10^{-1}$). For higher mass systems ($M_1 \geq 
0.8~M_{\odot}$), on the other hand, this condition is met at short periods even
 with $(1-\alpha) = (0.1\div 1)$.\\
The choice of low magnetic moments is completely arbitrary here and justified
only because it gives ultrashort period systems where $\dot{\omega}_o$ is 
determined by GW alone. 
In paper II we apply the model to RX J0806+15 and RX J1914+24, showing that 
such low magnetic moments are not simply convenient but seem to be 
required for the model to work, which it does quite well indeed.\\
In summary, B can be neglected even for highly asynchronous systems at orbital 
periods longer than a few thousands seconds and $\mu_1 < 10^{31}$ G cm$^3$, 
while at shorter periods $(100 \div 2000)$ s this approximation holds only if 
they are almost synchronous ($1-\alpha  < 10^{-1}$) or have sufficiently 
high-mass primaries.
In Fig. \ref{period} 
%and Fig. \ref{alfa} 
we show an example of the numerical integration of the evolutionary equation 
(\ref{alfaevolution}) for two different values of the primary magnetic moment.
Representative, although arbitrary, system parameters and initial conditions 
have been chosen (see captions). The early phase of orbital spin-down (the 
period actually changes just slightly) for the highly magnetized system is seen
 clearly in the upper curve of Fig. \ref{period}.
\subsection{A Steady-State Solution}
\label{steadystate}
At this point we introduce a major simplification of the problem by assuming
that the timescale over which $\alpha$ changes is considerably shorter than 
the evolutionary timescale of $\omega_o$. \\
On one hand, this restricts the physical regime of interest to conditions that
will be carefully explored in the next section. On the other hand it affords a 
very simple and straightforward solution, providing physical insight on the 
problem at hand.
An exact, non-linear and time-dependent study will be carried out in a future 
investigation (Dall'Osso et al. in preparation).\\
With the above approximations, eq. (\ref{alfaevolution}) becomes a first order 
differential equation for $\alpha$ with constant coefficients whose solution 
is given by:
\begin{equation}
\label{alfasolution}
\alpha(t) = (\alpha_0 - \alpha_{\infty}) \mbox{e}^{-t/ \tau_{\alpha}} + 
\alpha_{\infty}
\end{equation}
where $\alpha_0$ is the (arbitrary) initial value and $\tau_{\alpha}$ and 
$\alpha_{\infty}$ are defined as follows:
\begin{eqnarray}
\label{deftauealfainfty}
\tau_{\alpha} & = & \left(\frac{\dot{\omega}_o}{\omega_o} 
+ \frac{k}{I_1} \omega^{11/3}_o\right)^{-1} \nonumber \\
\alpha_{\infty} & = & \frac{k \omega^{11/3}_o}{I_1} \tau_{\alpha}~~\mbox{ 
hence }~~ \lim_{t \rightarrow \infty} \alpha = \alpha_{\infty} < 1
\end{eqnarray}
This solution requires $\dot{\omega}_o$ to be driven by GW alone and both 
($\dot{\omega}_o/\omega_o$) and $\omega_o$ to be strictly constant: the latter 
requires $\tau_{\alpha} \ll \tau_o$, as both the above quantities evolve on 
a timescale $\sim \tau_o$.\\
The timescale $\tau_{\alpha}$ and the parameter $(1- \alpha_{\infty})$ 
themselves are determined by the orbital period and its derivative, so they 
will be subject to secular evolution as well: in particular, they both 
decrease as the system shrinks. Hence, once the system has reached the 
asymptotic asynchronous state - starting from an arbitrarily asynchronous 
configuration - $\alpha$ will be locked to its ``local'' steady-state value 
during the subsequent evolution.\\
Expressions for $\tau_{\alpha}$ and W can now be rewritten in terms of 
$\alpha_{\infty}$, $I_1$ and the measured quantities $\omega_o$ and 
$\dot{\omega}_o$. 
%Indeed $\alpha_{\infty} =  A/(N+A)$ and, with 
%some algebra, it obtains:
%
\begin{eqnarray}
\label{quantitiesdefined}
\tau_{\alpha} & = & \frac{\omega_o}{\dot{\omega}_o} (1-\alpha_{\infty}) =
\tau_o (1-\alpha_{\infty}) \nonumber \\
W & = & \alpha_{\infty} I_1 \omega_o \dot{\omega}_o \frac{(1-\alpha)^2}
{1- \alpha_{\infty}}
\end{eqnarray}
\subsection{Validity of the exponential solution}
\label{limval}
\begin{figure}[h]
\includegraphics{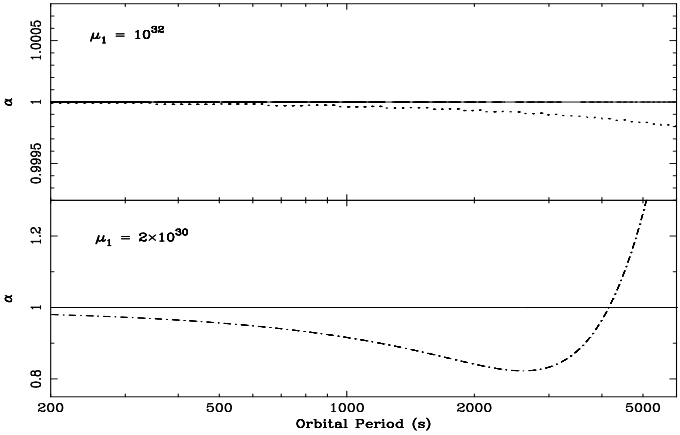}
\vspace{5.4cm}
\caption{Late evolution of the asynchronism parameter $\alpha$, obtained 
from the numerical integration of eq. (\ref{alfaevolution}), for a Double
Degenerate Binary with $M_1 = 1~M_{\odot}$, q=0.2, and two different values
of $\mu_1$. These are indicated in the figure and expressed in G cm$^3$. The 
initial spin period is $\sim$ 4 hrs and $\alpha_{in}=$ 10 but, once $\alpha = 
\alpha_{\infty}$, its value is independent of the previous history of the 
system: it is determined only by system parameters, given the orbital period. 
The dependence of $\alpha_{\infty}$ on $\omega_o$ can be understood here only 
at a qualitative level; in $\S$\ref{energy} this we obtain it in full 
generality.}
\label{alfa}
\end{figure}

The picture introduced in the previous section holds under well defined
assumptions that will be addressed here. 
The condition that $\alpha$ changes rapidly with respect to the timescale 
over which $\omega^{11/3}_o$ evolves corresponds to the following statement: 
after a time $\tau_{\alpha}$, $\omega^{11/3}_o$ must have changed by a small
amount, say less than 10\%, in order for its approximation to a constant 
coefficient to be acceptable. Eq. (\ref{quantitiesdefined}) then implies:
\begin{equation}
\label{limitofvalidity1}
1 - \alpha_{\infty}  <  \frac{3}{110}
\end{equation}
that can be translated to the following condition on the orbital period:
\begin{eqnarray}
\label{limitofvalidity2}
\omega_o > \omega_{fast} & = & \frac{(36 I_1)^{\frac{3}{11}}~ 
[(H/ \Delta d)\jmath(e)]^{\frac{3}{11}}~[G M_1 (1+q)]^{\frac{1}{2}}}
{(2 \sigma R^3_2 R^{3/2}_1)^{\frac{3}{11}}} \nonumber \\
 & ~ & \times \left(\frac{c}{\mu_1}\right)^{\frac{6}{11}} 
\left(\frac{\dot{\omega}_o} {\omega_o}\right)^{\frac{3}{11}}
\end{eqnarray}
We insert in this equation the pure GW expression for $\dot{\omega}_o
/\omega_o$ and, for the sake of simplicity, write it as $B \omega^{8/3}_o$,
where the constant $B$ incorporates all physical constants and system constant
parameters. It readily obtains:
\begin{equation}
\label{omegatilde}
\omega_o > \omega_{fast} = 36 \frac{B I_1}{k}
\end{equation}
which translates to the following limiting period:
\begin{eqnarray}
\label{limitofvalidity3}
P & < & P_{fast} = 1.2 \times 10^3 \left(\frac{M_1}{0.9~M_{\odot}}
\right)^{-\frac{7}{2}} \left(\frac{\mu_1}{2.5 \times 10^{30}}\right)^2
\nonumber \\
 & & \left(\frac{R_1} {6 \times 10^8}\right)^{\frac{3}{2}} \times  
\left(\frac{R_2}{1.7 \times 10^9}\right)^3   
\frac{\left(\frac{I_1}{2.8 \times 10^{50}}\right)^{-1}}{q (1+q)^{\frac{3}{2}} 
\left(\frac{H} {\Delta d}\right)}
\end{eqnarray}
where c.g.s. units in the normalizations and $\jmath(e)$ in the denominator 
have been omitted.
\begin{figure}[h]
\includegraphics{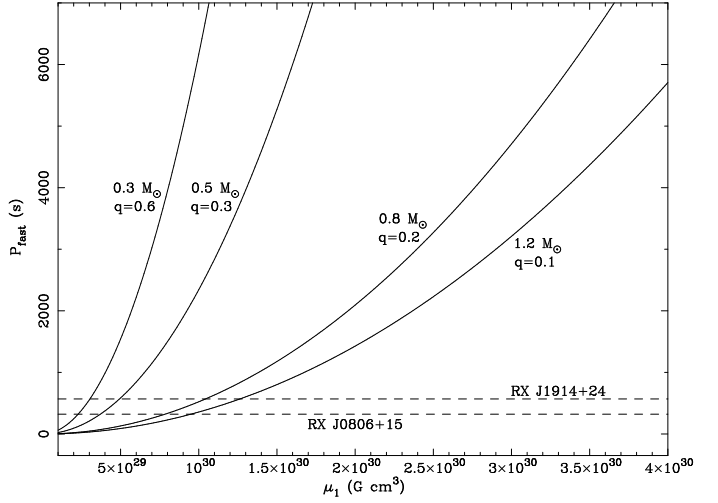}
\vspace{6.8cm}
\caption{The orbital period for which $\tau_{\alpha} \ll \tau_o$ 
($P_{fast}$) as a function of the primary magnetic moment ($\mu_1$) for four 
different values of $M_1$. The measured $\dot{\omega}_o$ and 
$\omega_o$ of both candidate ultrashort period binaries - RX J0806+15 and RX 
J1914+24 - constrain secondary masses in the range ($0.1 \div 0.35$) M$_
{\odot}$ approximately, increasing for decreasing $M_1$ (see paper II): we 
chose here $q$ in order to give an approximately constant $M_2 \sim 0.2$ 
M$_{\odot}$, for illustrative purposes. The rightmost curve has a significantly
 lower secondary mass, reflecting a similar result of paper II, although it 
would have changed very little if we had allowed for a somewhat larger $M_2$.
A general trend is clear: if $\mu_1 > 4 \times 10^{30}$ G cm$^3$, the 
condition for fast synchronization is reached for fairly long orbital periods, 
whatever the system masses. Further, the value of $P_{fast}$ is sensitive to 
the primary mass, low-mass systems reaching fast synchronization at longer 
periods than high mass ones for this particular choice (and most plausible 
choices) of the mass ratios. Finally, the measured orbital periods of the 
candidate ultrashort period binaries are indicated by the dashed lines: if 
their measured orbital spin-up is due to GW alone, they should both be in the 
fast synchronization regime (or close to it), unless $\mu_1 < 10^{30}$ G 
cm$^3$. This is discussed in paper II and is reported here just for the sake of
 illustration.}
\label{pmin}
\end{figure}
From the above one obtains fundamental indications concerning the 
evolutionary scenario implied by the UIM: there exists a critical period
(in the time-dependent case this will essentially correspond to a range of 
periods) shortwards of which the synchronization becomes fast compared
to the system orbital evolution. Double degenerate binaries are thought to be 
born with orbital periods of a few hours: hence, if asynchronous at 
birth, their asynchronism will not change much faster than $\omega_o$ until 
they shrink to a sufficiently small orbital separation.
All that happens \textit{before} a system meets the fast synchronization
requirement depends on its essentially unknown initial conditions (initial 
value of $\alpha$) and on the particular evolutionary path it follows. However,
once the orbital period is short enough, the UI mechanism becomes so efficient
as to cause fast synchronization of the primary spin, on a timescale 
$\tau_{\alpha} \ll \tau_o$: in this regime, the value of $\alpha$ is brought 
to the corresponding $\alpha_{\infty}$ while $\omega_o$ remains essentially 
unchanged. \\
The state of a system \textit{after} it has gone through fast synchronization 
is completely independent of initial conditions and its previous evolution; it
becomes a function of the orbital period and the fundamental parameters
$M_1, q~\mbox{and}~\mu_1$ only.\\
In particular, higher mass systems reach the fast synchronization regime at 
shorter periods than lower mass ones (see Fig. \ref{pmin}), because in the 
latter the current dissipation rate is stronger, for a given orbital period 
and magnetic moment and, at the same time, GW are weaker. Hence, high mass 
sytems can reach shorter periods still maintaining a relatively high degree of 
asynchronism, whose exact value depends on the initial one.\\
Upon inverting relation (\ref{limitofvalidity3}) (or directly from the example
of Fig. \ref{pmin}) it is also found that, if $\mu_1 > 10^{31}$ G cm$^3$, 
the fast synchronization regime is reached at periods longer than $10^3 
\div 10^4$ s, whatever the component masses. Given the strong efficiency of UI 
in such highly magnetic systems, they will be characterized by very low steady 
state values of the asynchronism and very short timescales to reach it. 
Therefore, somewhat contrary to intuition, the luminosity of highly magnetized 
systems at short orbital periods will be quite low because of the tiny degree 
of asynchronism they can sustain.\\
Finally, recall that we have found in $\S$ \ref{steadystate} that, at periods 
less than 
$\sim$ 2000 s, UI is negligible with respect to GW in low-mass sytems only if 
their asynchronism is quite low. High-mass sytems, on the other hand, may 
fulfil that requirement even with a higher degree of asynchronism. However, 
from eq. (\ref{limitofvalidity3}) it follows that low-mass systems reach 
steady-state at orbital periods significantly longer than $10^3$ s, so they 
will certainly have a low value of $(1-\alpha)$ shortwards of that. On 
the contrary, systems with sufficiently high-mass primaries (and low $\mu_1$) 
can reach periods $\sim 10^3$ s or less without experiencing fast 
synchronization and are ultimately more likely not to fulfil the requirement UI
 $<<$ GW at short orbital periods.
In Fig. \ref{alfa} we show 
the late evolution of the asynchronism parameter as obtained through a 
numerical integration of eq. (\ref{alfaevolution}). The same system parameters 
of Fig. \ref{period} have been used. The behaviour of $\alpha$ in the plot 
confirms that the system never becomes exactly synchronous as it shrinks;
the value of $\alpha_{\infty}$ has a marked dependence on $\mu_1$, in the same 
way expressed by eq. (\ref{deftauealfainfty}).
\section{Energy Budget in the UIM: powering the circuit through GW emission}
\label{energy}
There is a general line of reasoning, based on energetic arguments, that leads 
to the definition of a system's steady-state without solving the evolutionary
equation and without simplifying assumptions\footnote{In the following
 we neglect the small term $h(a)$ in $g(\omega_o)$ ($\S$\ref{general}). This 
term accounts for the secondary spin being tidally locked to the orbit and can 
be easily added to all formulas. It is neglected only for the sake of clarity, 
as is it does not alter the physical content of the discussion, leaving all 
emphasis on physically more relevant features and somewhat simplifying the 
notation.}.\\
In the UIM a simple electric circuit is devised, with the secondary star
acting as a generator and the primary as a resistance. The e.m.f. driving the
current is due to the asynchonism between the primary spin and the orbital 
motion: hence, we can write the total energy available to the circuit simply 
as:
\begin{equation}
\label{Euim}
E_{UIM} = \frac{1}{2} I_1 (\omega^2_1 - \omega^2_o) = E^{sync}_1 (\alpha^2 -1)
\end{equation}
where $E^{sync}_1 = (1/2) I_1 \omega^2_o$ is the primary spin energy for
perfectly synchronous rotation.\\
With this definition $E_{UIM} >0$ if $\omega_1 > \omega_o$ while it is
negative in the opposite situation. However, what counts is the absolute value
of $E_{UIM}$, so that a more negative value corresponds to a larger energy 
reservoir.\\
From the above equation, the rate of change of the total available energy is 
readily obtained:
\begin{equation}
\label{edot}
\dot{E}_{UIM} = I_1 \omega_o (\dot{\omega}_1 \alpha - \dot{\omega}_o)
\end{equation}
and we see that when $\alpha =1$ the available electric energy goes to zero 
but has a nonzero, negative first derivative; indeed, in this situation 
$\dot{\omega}_1 =0$ because the circuit is switched off, while GW still 
contribute to $\dot{\omega}_o$. Because of this, $\alpha$ becomes smaller than 
1 and the negative $\dot{E}_{UIM}$ - that with $\alpha >1$ implied a decreasing
energy reservoir - now causes the circuit energy to increase again, in absolute
 value: GW start feeding energy to the circuit from this point on.\\ 
In the UIM, dissipation of the electric current works against the e.m.f. and
acts to synchronize the primary spin with the orbital motion. This is done
through the torque $N_1 = I_1 \dot{\omega}_1$, that transfers angular momentum
between the primary and the orbit. The torque is readily obtained remembering
that the primary spin derivative is (see eq. \ref{alfaevolution}):
\begin{equation}
\label{spindot}
\frac{\dot{\omega}_1}{\omega_1} = \frac{W}{\alpha (1-\alpha) I_1 \omega^2_o}
\end{equation}
The rate of work done by the torque on the primary - the rate of spin energy 
change $\dot{E}^{(1)}_{spin}$ - is just $N_1 \omega_1$ so that, finally, we 
have:
\begin{equation}
\label{spindownenergy}
\dot{E}^{(1)}_{spin} = \frac{\alpha}{1-\alpha} W
\end{equation}
Since no net angular momentum loss arises from the coupling, the same torque 
with an opposite sign acts on the orbit as well, its rate of work being now 
$\dot{E}^{(UIM)}_{orb} = - N_1 \omega_o = -\dot{E}^{(1)}_{spin} / \alpha = 
-W/(1-\alpha)$.\\
Eq. (\ref{spindot}) and (\ref{omegadot}) can now be subsituted into eq. 
(\ref{edot}) and, remembering eq. (\ref{spindownenergy}), the following is 
obtained:
\begin{equation}
\label{dissipationrate}
\dot{E}_{UIM} = \dot{E}^{(1)}_{spin} - 3 \frac{E^{sync}_1}{E_g} \dot{E}_g
 - 3 \dot{E}^{(UIM)}_{orb} \frac{E^{sync}_1}{E_g}
\end{equation}
This last expression shows that the rate of change of $E_{UIM}$ receives three 
different contibutions: 
\begin{itemize}

\item the first term, $\dot{E}^{(1)}_{spin}$, represents the rate of change of 
the primary spin because of the work done by the Lorentz torque. Obviously, 
this term always causes the absolute value of $E_{UIM}$ to decrease, since 
$E_{UIM}$ and $\dot{E}^{(1)}_{spin}$ always have opposite signs. 

\item the second term is always \textit{negative} and represents the effect of
GW emission. Having a negative sign it causes the absolute value of $E_{UIM}$ 
to decrease if $\omega_1 > \omega_o$, but in the opposite case it causes 
$|E_{UIM}|$ to increase and, thus, represents a mechanism for injecting energy 
in the circuit.\\
This is the only term that would be present if there was no spin-orbit 
coupling at all. Its effect is better understood in this framework, indeed. 
Suppose $\omega_1$ is a constant, greater than $\omega_o$: since GW cause 
$\omega_o$ to increase, the ratio $\alpha$ will decrease, thus consuming the 
available electric energy. From the point when $\alpha=1$ on, the further 
increase of $\omega_o$ makes $\alpha$ to become increasingly smaller than 1, 
thus powering the circuit battery once again: while the loss of orbital energy 
consumes the electric energy when $\alpha>1$, it powers the electric circuit 
when $\alpha <1$, thus showing that the sign of $E_{UIM}$ in our definition 
(eq. \ref{Euim}) has a direct physical meaning. 

\item the last term describes spin-orbit coupling itself. It represents
the rate of work done on the orbit by the torque $N_1$, times the ratio 
between the primary and orbital moments of inertia, where the latter is 
defined as $I_o = M_1 a^2 q/(1+q)$ and $E_g = -(1/2) I_o \omega^2_o$.\\
This term always acts to increase $E_{UIM}$ \textit{in absolute value}, as it 
can be checked (remember that $E_g <0$ by definition). 

\end{itemize}
\subsection{Conserving the Total Electric Energy}
\label{conserving}
The above considerations lead naturally to the following question: given that
only one term out of three in eq. (\ref{dissipationrate}) always dissipates 
electric energy, can a condition be reached where dissipation is balanced by 
the terms feeding energy to the circuit? If this was the case, one would 
expect the flow of currents never to stop and the X-ray emission associated to 
their dissipation never to fade away.\\
To answer this point consider again $\dot{E}_{UIM}$ in the form of eq. 
(\ref{edot}): we see that the energy first derivative is zero if 
$\dot{\omega}_1 \alpha = \dot{\omega}_o$. Further, we can write the second 
derivative:
\begin{equation}
\label{eduedot}
\ddot{E}_{UIM} = I_1 \dot{\omega}_o (\dot{\omega}_1 \alpha - \dot{\omega}_o)
+ I_1 \omega_o (\ddot{\omega}_1 \alpha + \dot{\omega}_1 \dot{\alpha}
- \ddot{\omega}_o)
\end{equation}
and see that this is zero as well, if $\dot{\omega}_1 \alpha = \dot{\omega}_o$.
Hence, if this equality is verified, it implies that the available energy in 
the generator stays constant following the system evolution: the dissipation 
of electric current is exactly balanced by the spin-orbit coupling and GW 
emission. We call this the ``energy steady-state''.\\
We remember here that in the present work we are essentially concerned with 
the case when $\dot{\omega}_o > 0$. 
In this regime, it must be $\dot{\omega}_1 \alpha > 0$ in order for $\alpha 
\dot{\omega}_1 = \dot{\omega}_o$ to be possible. When the primary spin has the 
same verse as the orbital motion ($\alpha >0$) this condition exists only for 
$\alpha <1$, because the primary spins down ($\dot{\omega}_1 < 0$) if 
$\alpha>1$. When the primary spin is antialigned ($\alpha<0$), 
eq. (\ref{spindot}) shows that $\dot{\omega}_1$ is always positive and again 
the condition $\dot{E}_{UIM} =0$ cannot be met.\\
Therefore, the energy steady-state \textit{in the presence of orbital spin-up} 
exists only with the primary spin aligned with and somewhat slower than the 
orbital motion. This is not surprising, since we have seen that GW feed energy 
to the circuit only when $\alpha <1$.\\
When $\alpha >1$ and $\dot{\omega}_o >0$, the energy first derivative receives 
a negative contribution from both terms in eq. (\ref{edot}) and $E_{UIM}$
is an ever decreasing quantity. A steady-state solution with $\alpha>1$ can in 
principle exist only if $\dot{\omega}_o < 0$, \textit{i.e.} when
spin-orbit coupling dominates over GW emission. If sufficiently strong, this
effect could overcome all others and allow $E_{UIM}$ to stay 
constant or even increase while the orbit widens. \\
However, $\alpha^{en}_{\infty} = |\dot{\omega}_o|/|\dot{\omega}_1| > 
1$ must be compared to the definition of the two derivatives (eq. 
\ref{omegadot} and eq. \ref{spindot}). Consider the largest possible value of 
$\dot{\omega}_o$, the one obtained if GWs are completely negligible; even in 
this case, in order for $\dot{\omega}_o > \dot{\omega}_1$ to hold we must have 
$\alpha < 3 (I_1/I_o)$. 
The right-hand side of this disequality is very hardly larger than 1 (one needs
 quite small component masses - $M_1 \leq 0.5 M_{\odot}), q\leq 0.2$ - and an 
orbital period shorter than 400 s. We conclude that steady-state during 
spin-down is very unlikley to exist in real systems.
Furthermore, this case is of no relevance for the two ultrashort period 
binaries RX J0806+15 and RX J1914+24 and we will neglect it from here on. For 
the sake of completeness, in Appendix B we show that orbital spin-down in the
UIM is necessarily a transient phase, after which every system must eventually
start shrinking.
\subsection{The ``Energy Steady-State'' with orbital spin-up}
\label{constant}
We can now write $\dot{\omega}_1$ and $\dot{\omega}_o$ explicitly in 
eq. (\ref{edot}) taking their expressions given, respectively, in eq. 
(\ref{alfaevolution}) and eq. (\ref{omegadot}). From this we find a simple 
expression for the value of $\alpha$ at which $E_{UIM}$ stays constant,
that we indicate as $\alpha^{en}_{\infty}$ to distinguish it from the 
steady-state $\alpha_{\infty}$ of the approximate solution ($\S 
\ref{steadystate}$).\\
By imposing $\dot{E}_{UIM} =0$ it obtains, after a little algebra\footnote{We 
leave $(\dot{\omega}_o/\omega_o)$ because this is a measured quantity,
independent of assumed system parameters. Writing $\dot{\omega}_o$ 
explicitly, $\alpha_{\infty}$ is obtained as a function of system 
parameters only. Put
$B = (3/2E_g) I_1 \omega^2_o$ and $C = (3/2) (\dot{E}_g/E_g) 
I_1/(k \omega^{11/3}_o)$.\\ Then $\alpha_{\infty} = (1/2) \left[(1-B) \pm 
\sqrt{(1+B)^2 - 4C}\right]$. Requiring the expression under square 
root to be positive, a limit very similar to eq. (\ref{omegamin}) is 
obtained, coincident with it if $I_1/I_o \rightarrow 0$, an appropriate limit
indeed.}
%again we see that $(1-\alpha)> $0 is required since the right-hand term is
%positive}:
%
\begin{equation}
\label{alfasteady}
\alpha^{en}_{\infty} (1-\alpha^{en}_{\infty}) = \frac{I_1~(\dot{\omega}_o / 
\omega_o)}{k \omega^{11/3}_o} = A
\end{equation}
a second order equation with the following roots:
\begin{equation}
\label{roots}
\alpha^{en}_{\infty} = \frac{ 1 \pm \sqrt{1-4 A}}{2}
\end{equation}
From the above we see that, in order for a steady-state solution to
exist, it must be $A \leq 1/4$. If $A$ is greater than that, the discriminant 
of the above equation is negative, which ultimately implies $\dot{E}_{UIM} < 0 
$.\\
We address the meaning of the requirement $A<1/4$ starting from the definition:
\begin{equation}
\label{Asmallerthan}
\frac{\dot{\omega}_o}{\omega_o} \leq \frac{1}{4} \frac{k \omega^{11/3}_o}{I_1}
\end{equation}
that, remembering the general expression (\ref{omegadot}) for $\dot{\omega}_o
/ \omega_o$ and writing the pure GW contribution to it as $B \omega^{8/3}_o$, 
leads to the following:
\begin{eqnarray}
\label{Asmallerthanbis}
B \omega^{\frac{8}{3}}_o & - & \frac{3 k \omega^{\frac{17}{3}}_o}{2 E_g} (1-\alpha) \leq \frac{1}{4} \frac{k \omega^{\frac{11}{3}}_o}{I_1}~~~\mbox{ or} 
\nonumber \\
1-\alpha & \leq & \frac{2 E_g B}{3 k \omega^3_o} - \frac{1}{6} \frac{E_g}
{I_1 \omega^2_o} = n(\omega_o)
\end{eqnarray}
which assures that $A<1/4$. \\
This condition can in principle be met for any value of $\alpha$,
greater or smaller than 1, depending on system parameters and initial 
conditions, because it concerns only the existence of a steady-state 
configuration but not the fact that the system has actually reached it. 
However, the condition must clearly hold when the system 
actually reaches steady-state, and thus has $\alpha = 
\alpha^{en}_{\infty} < 1$: this implies $n(\omega_o)>0$, since $1-
\alpha^{en}_{\infty}$ is positive, which translates to:
\begin{equation}
\label{omegamin}
\omega_o > 4 \frac{B I_1}{k} = \frac{\omega_{fast}}{9} 
\end{equation}
This corresponds to a limiting period $P_{steady} = 9 
P_{fast}$ (eq. \ref{limitofvalidity3}) longwards of which
the condition $A<1/4$ for the existence of an ``energy steady-state'' is 
incompatible with $\alpha^{en}_{\infty} <1$. Stated differently, this result 
implies than no steady state exists at orbital periods 
longer than $P_{steady}$ and $\dot{E}_{UIM}$ can only be negative in these 
cases.\\
\subsection{Asynchronous steady-state: the general expression and its 
approximation}
\label{comparison}
We comment here on the relation between the ``energy steady-state'', obtained 
in full generality, and the one derived in the previous section with neglect 
of the spin-orbit coupling and temporal variations of $\omega_o$.\\
First of all, the ``energy steady-state'' has two different roots while in 
$\S$ \ref{steadystate} we only found one. This is due to the non-linearity of 
the problem: the complete evolutionary equation for $\alpha$ has a second order
term, which leads to the two roots of eq. (\ref{roots}). On the other hand, in 
$\S$ \ref{steadystate} we explicitly reduced the equation to the linear form, 
thus excluding one of the two solutions. Concerning this point we can note 
that, once $A<1/4$, one solution tends to be close to 1 (although always 
somewhat smaller) and the other tends to zero (although always somewhat 
higher). Therefore, it seems natural to neglect the latter if the system 
evolves from $\alpha>1$ towards steady-state. If the system started from an 
almost zero initial spin, on the other hand, the solution with the smallest 
$\alpha^{en}_{\infty}$ would be met first and should be taken into account.\\
It can be easily realized that the approximate expression for $\alpha_{\infty}$
 (eq. \ref{deftauealfainfty}) differs from the general one for 
$\alpha^{en}_{\infty}$ (eq. \ref{roots}) just by a small quantity. Indeed, 
from eq. (\ref{deftauealfainfty}) we can write:
\begin{equation}
\label{riscritta}
1-\alpha_{\infty} = \frac{A}{1+A} 
\end{equation}
from which the fast synchronization condition (eq. \ref{limitofvalidity1}) 
becomes:
\begin{eqnarray}
\label{smallA}
\frac{A}{1+A} & \leq & \frac{3}{110}~~\mbox{ or} \nonumber \\
A & \leq & \frac{3}{107}
\end{eqnarray}
In this case, since from eq. (\ref{deftauealfainfty}) $\alpha_{\infty} = 
(1+A)^{-1}$ and $A$ must be this small, we can re-write it, at first order in 
$A$, as $\alpha_{\infty} \simeq 1-A$. Expanding eq. (\ref{roots}) at first 
order in $A$ as well, the same expression is obtained, thus proving that
$\alpha^{en}_{\infty}$ and $\alpha_{\infty}$ are coincident, at lowest order 
in $A$. \\
Further, eq. (\ref{smallA}) shows that an upper limit on $A$ is not different 
from a constraint on the synchronization timescale, so that condition $A<1/4$
can be given this more intuitive meaning, resulting simply in a less 
restrictive statement than the short synchronization condition (eq. 
\ref{limitofvalidity1}). It happens to be less restrictive than that just 
because it describes a general result, based on no approximation. 
$A< 3/107$ corresponds indeed to a very specific case, where 
temporal variations of $\omega_o$ are so slow that can be neglected in the 
evolution of $\alpha$ (although they ultimately determine the very existence 
of an asynchronous steady-state). The result of this section does not
rely on that assumption: it adds to the previous analysis the 
conclusion that a well defined steady-state exists even when 
temporal variations of $\omega_o$ are non-negligible. Nevertheless, the 
details of if and how the system evolves towards it cannot be addressed in the 
time-independent, linear approximation, in the general case.
Hence, steady-state is already defined for systems meeting the 
requirement (\ref{omegamin}), but these are evolving towards it on a
 timescale that is not much shorter than $\tau_o$: the fast synchronization 
regime ensues only at the even shorter orbital period $P_{fast}$.\\
The discussion of this section shows that \textit{the steady-state degree of 
asynchronism towards which the system evolves corresponds to a stationary state
for the available electric energy}: the system adjusts its parameters as to
dissipate only the energy that is fed by GW emission and spin-orbit coupling, 
maintaining a somewhat negative, constant electric energy reservoir. 
The latter can be expressed as $E_K = - E^{sync}_1 (1 - A +\alpha^{en}_{\infty}
) > - E^{sync}_1 (2-A) $.
\section{Conclusions}
\label{conclusions}
In the present work we have discussed some important implications of the 
Unipolar Inductor Model applied to ultrashort period DDBs, that had been 
overlooked in previous works. In particular, we have focussed attention on 
systems with orbital spin-up. The main result of our study can be summarized 
as follows:
%\begin{itemize}
\textit{in the framework of the UIM, and in systems whose orbit is shrinking,
 the dissipation of the e.m.f. and associated currents can be balanced by 
spin-orbit coupling and mainly, when the primary spin is slower than the 
orbital motion, by the emission of gravitational waves}. \\
When the asynchronism parameter $\alpha$ is smaller than unity GW feed the 
the circuit battery - at the rate at which the orbit shrinks - by driving the 
primary spin out of synchronism and through the orbital spin up itself. 
This is a remarkable circuit, in which gravitational energy is 
``converted'' to electric energy, powering a continuous flow of currents. An 
equilibrium with the energy dissipation process can thus be reached, such that 
the electric circuit is not expected to switch off.\\
In the early evolutionary stages - long orbital periods, say a few hours - it 
is likely that the primary spin be faster than the very slow orbital motion: 
the degree of asynchronism decreases at a rate comparable to that at which 
$\omega_o$ itself evolves, because the Lorentz torque is initially much weaker 
than the GW torque. \\ 
When the orbital period becomes sufficiently short ($P <P_{steady}$), on the 
other hand, a steady-state solution exists independent of the current value of 
$\alpha$ and such systems start evolving towards it. When they reach the even 
shorter period $P_{fast}$ they enter the fast synchronization regime, where 
steady-state is achieved in a very short time compared to $\tau_o$. The 
existence of this regime is due to the fact that the rate of dissipation of 
currents is a stronger function of $\omega_o$ than GW are: hence, a point is 
reached where most of the residual electric energy is consumed over a short 
time - during which GW affect it only very slightly - at the expense of the 
primary spin. 
From this point on the circuit is forced to work in an \textit{almost} 
synchronous state, in which the associated dissipation rate is balanced by the 
energy being fed by GW and spin-orbit coupling.\\
Therefore, one may say that the electric energy reservoir of systems born with 
$\alpha>1$ (and sufficiently weak primary magnetic moment to allow the orbit to
shrink) must initially decrease. 
They dissipate this energy at a rate determined by the orbital parameters, 
magnetic moment \textit{and} degree of asynchronism. 
Their lifetime is thus a dependent variable, fixed by the ratio between the 
initial $E_{UIM}$ and the dissipation rate $\dot{E}_{UIM}$. When they reach 
$\alpha=1$ the initial reservoir is completely consumed.\\
At this point a change in the nature of the circuit occurs: GW revive it by 
substracting further orbital energy (GW ``inject'' negative energy in the 
circuit) and $\omega_o$ continues increasing, thus becoming larger than 
$\omega_1$. As this happens, the primary spin starts tracking the orbital 
spin-up because of the spin-orbit coupling, although remaining somewhat lower
than $\omega_o$. Since $\alpha <1$, the steady-state condition $\dot{\omega}_1 
= \dot{\omega}_o/\alpha$ can be met, \textit{if} the orbital period has 
become sufficiently short (eq. \ref{omegamin}). Once steady-state is reached, 
$\omega_1$ continues increasing somewhat more rapidly than $\omega_o$: 
$\alpha$ becomes a slowly incresing parameter, approaching 1 from below over a 
timescale much longer than $\tau_o$ itself. However, it cannot reach exactly 
unity since the total energy in the circuit remains constant and cannot 
vanish.\\
The system lifetime is now virtually infinite (apart from the possible onset 
of mass transfer) while the degree of asynchronism becomes a dependent 
variable, being fixed by the condition that the dissipation rate always match 
the rate of work done by GW and spin-orbit coupling.\\
Several details of the UIM deserve further investigation in order to better
assess its predictions; a number of observational constraints that were not 
considered here should still be taken into account.
In particular, the simple UIM we have appealed to has been debated since its 
early proposal for the Jupiter-Io system. Assuming a perfectly field-aligned 
current system implies complete neglect of plasma inertia effects. 
In fact, it has been argued by several authors (Drell et al. 1965, Neubauer 
1980, Neubauer 1998, Russell \& Huddleston 2000, Saur et al. 2004, Lopes
\& Williams 2005) that ``field-aligned'' currents could be associated to 
standing (in Io's frame) Alfven disturbances; these distort significantly 
the dipole (unperturbed) field lines. The very presence of magnetic interaction
 between Jupiter and its satellite, and the presence of a large-scale 
current flow between them, is not questioned. Therefore, overall system 
energetics, the intensity of the current flow and exchange of angular momentum 
through the currents themselves remain essentially valid. On the other hand, 
the way the coupling works may be more complicated than assumed here.  
Plasma inertia effects could, even significantly in principle, affect the 
efficiency of the coupling and the geometry of the current flow, as well as
its stability. \\
The main goal of the present work was that of demonstrating the viability of 
the model, by showing that the UI phase is not short-lived because the emission
 of GW can keep the circuit working at any time. To this aim, we have referred 
to the basic UIM since this appeared to give the best physical insight into 
this problem. Account for MHD effects, which may well be of relevance to a more
detailed description of system properties, is delayed to future work.
\section{Appendix A}
\label{a}

Here we check that the primary tidal synchronization timescale is expected
to be longer than the GW evolutionary timescale of the binary system, in the 
weak viscosity approximation first introduced by Darwin (1879). 
In this approximation it is assumed that the star shape has a
quadrupole distortion (to leading order) induced by the tidal influence of the
companion. In a frame corotating with the star under study, the asynchronism 
between spin and orbit reflects in an apparent rotation of the companion at the
beat frequency $\omega_{b} = \omega_1 - \omega_o$. If stellar matter had no 
viscosity at all, its shape would 
instantaneously adjust to the present position of the companion and the tidal
bulge would be aligned with the line of the centres. A small but finite 
viscosity introduces a (small) lag in the reaction of the tidal bulge to the 
changing position of the companion, so the tides will lag behind the line of 
the centres by a small time lapse $\tau$ which measures the viscosity itself. 
The small viscosity approximation translates in the requirement that $\tau$ 
be sufficiently small, so that the stellar shape at time $t$ will be adjusted 
as to align the axis of its tidal bulge with the position the companion had at 
time $t-\tau$.\\
The expression for the evolution of a component spin due to the tidal 
interaction\footnote{we consider only the equilibrium tide, whose 
effects are expected in general to be much stronger than the dynamical tide 
(Zahn 1977)} with a companion in this framework was derived by Hut (1981) in
Appendix A and is:
\begin{equation}
\label{hut}
\frac{1}{T_t} = \left( \frac{\dot{\omega}_1}{\omega_1}\right)_t = 3 
\frac{k}{T} \frac{q^2}{r^2_g} \left(\frac{R}{a}\right)^6 \frac{1-\alpha}
{\alpha} 
\end{equation}
where in that equation we have assumed a circular orbit ($e=0$) with orbital 
and spin axes aligned ($i=0$). 
Here $k$ is the apsidal motion constant of the star ($\sim 0.12$ for a white 
dwarf, Verbunt and Hut 1983), $r_g$ is the star gyration radius, defined as 
$I = r_g M R^2$ and $T$ is a characteristic timescale of the tide, related to 
$\tau$ simply by (Hut 1981):
\begin{equation}
\label{T-tau}
T = \frac{R^3}{G M \tau}
\end{equation}
Note the strong dependence of $T_t$ on the ratio $(R/a)$ that measures how 
small is the star with respect to its Roche lobe, an intuitive result indeed.
Significantly different values of $T_t$ are thus expected for components with
significantly different masses. \\
The lag time $\tau$ can be related to the mean viscosity of the star through 
(Alexander 1973):
\begin{equation}
\label{alexander}
\overline{\mu} = \frac{75}{224 \pi} \frac{G M^2_1}{R^4} k \tau
\end{equation}
In order to neglect tidal interactions in the evolution of the primary spin
we must require that $T_t$ be considerably longer than the orbital 
evolutionary timescale (driven by GW). Hence:
\begin{equation}
\label{requirement}
3 \frac{k}{T} \frac{q^2}{r^2_g} \left(\frac{R}{a}\right)^6 \frac{|1 - \alpha|}
{\alpha} < \frac{96}{50}\frac{\left(GM_1\right)^{5/3}}{c^5} \frac{q 
\omega^{8/3}_o}{(1+q)^{1/3}} 
\end{equation}
where the right-hand expression corresponds to one tenth of $\dot{\omega}_o / 
\omega_o$ expected from GW emission.\\
With the above expression we finally obtain (rescaling the orbital period to
the shortest known, that of RX J0806+15):
\begin{equation}
\label{tide}
\overline{\mu} < 10^{14} \frac{(1+q)^{5/3}}{q} \left(
\frac{M}{M_\odot}\right)^{\frac{14}{3}} R^{-7}_9 \left(\frac{P}
{321}\right)^{4/3} \frac{\alpha}{|1-\alpha|}
\end{equation}
where P is expressed in seconds and $R_9$ is the star radius in units of 
$10^9$ cm. We need to introduce numbers in order to check the meaning of this 
condition: we consider the case of a relatively massive (say $M_1 > 0.8~
M_{\odot}$) primary star and a significantly less massive secondary (say, 
$M_2 < 0.25~M_{\odot}$), in order to describe the more likley situation for 
the two candidate ultrashort period binaries (see paper II). From eq. 
(\ref{tide}) it is seen that the most favourable case for tidal 
synchronization to be efficient is that of short period systems, so we 
consider only them here.\\
With a primary mass $M_1 = 0.8~M_{\odot}$ it obtains, at $P \sim 1000$ s and 
$\alpha =2$ (an extremely asynchronous system at such a short period), 
$\overline{\mu} > 10^{17}$ g cm$^{-1}$ s$^{-1}$ in order for $T_t < 10 
\tau_o$, or $\overline{\mu} = 10^{18}$ for the two times to be comparable. In 
a more likely situation, such as the 321.5 orbital period source RX J0806+15, 
with $M_1 = 0.8~M_{\odot}$ and a smaller asynchronism, $(1-\alpha)
\sim 10^{-2}$, an even larger value for the mean viscosity obtains, 
$\overline{\mu} > 10^{18}$ g cm$^{-1}$ s$^{-1}$, in order for tidal effects to 
act on a timescale shorter than 10 $\tau_o$. 
Note that even stronger constraints obtain when considering systems with a 
larger orbital separation than RX J0806+15.\\
Hence, tides are likely to always have long a synchronization timescale with 
respect to the orbital evolutionary timescale, unless $\overline{\mu} \geq 
(10^{18}\div 10^{19})$ g cm$^{-1}$ s$^{-1}$, a hardly plausible value. \\
Indeed, according to Kopal (1968), plasma (or radiative) viscosity in 
non-degenerate stars is at most $\sim (10^3 \div 10^4)$ g cm$^{-1}$ s$^{-1}$. 
In a degenerate star, however, it could well be of the same order of magnitude 
or somewhat stronger than turbulent viscosity in normal stars with convective 
envelopes, $\geq 10^{10} \div 10^{11}$ g cm$^{-1}$ s$^{-1}$ (see also 
Alexander 1973).
Essentially the same conclusions were reached by Iben, Tutukov and Fedorova 
(1998). They found that a mean viscosity much larger than $10^{13}$ g 
cm$^{-1}$ s$^{-1}$ would be required in order for tidal synchronization to be 
efficient over the GW timescale: they referred to the suggestion of Smarr \& 
Blandford (1976) in order for a relevant viscosity to be obtained, namely that
a magnetic field $> 10^4$ G with a proper orientation may strongly enhance a
WD viscosity. In particular, they speculate an extreme value $\sim 10^{18}$ g 
cm$^{-1}$ s$^{-1}$ can be reached, assuming an electrical conductivity 
$\sim 10^{19}$ e.s.u. and a sufficiently deep layer where dissipation of tidal 
energy takes place. We stress that this highly speculative suggestion is the 
only one made in the literature for such a high value white dwarf internal 
viscosity.\\
We focus now on the secondary component: consider the representative 
case of a secondary with $M_2 \sim  0.2 M_{\odot}$: hence, a 
plausible value $ \overline{\mu}  \sim (10^{10} \div 10^{12}) $ g cm$^{-1}$s
$^{-1}$ suffices to make tidal synchronization at least as fast as orbital 
evolution even for long orbital periods ($\sim 10^4$ s). 
Overall, then, it seems at least plausible that secondary stars are 
efficiently synchronized by tidal forces in these systems. However, given that
some room for doubt may be left on this subject, we want to stress here that 
the condition of tidal synchronization of the secondary enters our model mainly
through the function $h(a)$, of the order unity, defined in $\S$ \ref{general}.
\\
Hence, as long as the spin of the secondary does not signficantly alter the
e.m.f. calculated in $\S$ \ref{general}, the major consequence of a free 
secondary's spin in our model would be setting $h(a)=1$, not a significant 
change indeed.
The conclusion of this section can only be that tidal effects are at least 
extremely unlikely to be important in the evolution of the primary spin in 
the systems considered here. Althoguh large uncertainties still exist in the 
determination of white dwarf viscosities, all estimates point to it being not 
sufficiently strong to affect the primary spin over a timescale at least 
comparable to $\tau_o$. \\
We note however that, even if tides acted on the same timescale as GW emission
(they would be faster than GW only if $\overline{\mu} > 10^{18}$, even in 
the extremely favourable case of highly asynchronous systems at a period 
$\leq$ 1000 s) this would in any case make UIM the strongest spin-orbit 
coupling mechanism once the fast synchronization regime is reached. Hence,
even in the regime where tides may possibly become as fast as the orbital 
evolution ($P \leq 10^3$ s), UIM would be much more efficient in affecting the 
primary's spin.\\
Finally, low-mass secondaries are much more easily synchronized by tidal 
interactions and it seems that they are fully consistent with having a tidal 
synchronization timescale shorter than $\tau_o$ even at relatively long 
orbital periods.
%
%
%\section{Appendix B}
%\label{b}
%
\section{Appendix B}
\label{b}
We have shown in the text that the state with $\alpha^{en}_{\infty} >1$ 
cannot exist for orbital periods longer than a few hundred seconds or for
$M_1+M_2 > 0.6 M_{\odot}$ or so.
Further, we have briefly discussed systems undergoing orbital spin-down,
assuming that they must eventually stop and reverse their orbital evolution.
However, this latter conclusion has not been demonstrated and we address it
here.
Namely, we show that a maximum orbital separation (period) exists in this 
case, beyond which GWs take over spin-orbit coupling and the two stars 
can only spiral-in.
We begin writing the condition that the orbit keeps widening, that is 
$\dot{\omega}_o/\omega_o < 0 $ or $ g^{-1}(\omega_o)[\dot{E}_g - W/
(1-\alpha)] < 0$ at any orbital period. Put $g(\omega_o) = 
- \overline{g} \omega^{2/3}_o$, $\dot{E}_g = B_g \omega^{10/3}_o$ and 
thus obtain:
\begin{equation}
\label{spindown}
\frac{B_g}{\overline{g}} \omega^{\frac{8}{3}}_o +\frac{k}{\overline{g}} 
\omega^5_o (1-\alpha) \leq 0
\end{equation}
from which:
\begin{equation}
\label{necessary}
\alpha \geq 1 + \frac{B_g}{k} \omega^{\frac{7}{3}}_o
\end{equation}
%
%as a necessary condition for orbital spin-down to continue despite GW 
%emission.\\
By adding or subtracting 1 from eq. (\ref{necessary}) we eventually obtain:
\begin{equation}
\label{obtained}
(\alpha^2 - 1) \geq \frac{2 B_g}{k \omega^{7/3}_o} \left(
1 + \frac{1}{2} \frac{B_g}{k \omega^{7/3}_o}\right)
\end{equation}
that would require an ever increasing $E_{UIM}$, in order for spin-down to
continue.\\ 
Concerning this point note that, \textit{during orbital spin-down}, this 
condition can be met if spin-orbit coupling transfers more energy to the orbit 
than that lost by dissipation of currents and by the decrease of $\omega_o$. 
Indeed, when $\alpha >1$ ($E_{UIM} > 0$) spin-orbit coupling is the only term 
with a positive contribution to $\dot{E}_{UIM}$ (eq. \ref{dissipationrate}). 
Hence, in order for $E_{UIM}$ to increase, remembering eq. 
(\ref{dissipationrate}) and the relation $ (\dot{E}^{(1)}_s/ \alpha) = 
-\dot{E}^{(UIM)}_{orb}$ (cfr. $\S$\ref{energy}), we obtain: 
\begin{equation}
\label{rewritten}
\alpha \leq 3 \frac{E^{sync}_1} {|E_g|} \left(1 - 3 \frac{|\dot{E}_g|}
{\dot{E}^{(UIM)}_{orb}}\right)
\end{equation}
where all terms have been written as to be positive. In our hypothesis 
spin-orbit coupling ($\dot{E}^{(UIM)}_{orb}$) is stronger than GWs 
($\dot{E}_g$): in particular, in the extreme case that GWs were completely 
negligible, one would obtain from eq. (\ref{rewritten}) the largest upper limit
 on $\alpha$. However, even the largest upper limit is too constraing: indeed, 
neglecting the second term on the righ-hand side of eq. (\ref{rewritten}):
\begin{equation}
\label{increase}
\alpha \leq 3 \frac{I_1}{I_o}
\end{equation}
the orbital momentum of inertia $I_o$ being defined in $\S$\ref{energy}.\\
As already stated in the text, the right-hand side of this equation is very 
hardly greater than 1, especially for periods longer than a few hundred 
seconds, while $\alpha >1$ is a necessary condition here. So, during orbital 
spin-down, $E_{UIM}$ is not expected to increase as the system evolves. Thus 
write:
\begin{equation}
\label{dependence}
\alpha^2 -1 \propto \omega^{-(2-\delta)}_o
\end{equation}
where $\delta$ is positive and can be smaller or larger than 2. \\
The evolution of the asynchronism is thus not fast enough for $\dot{\omega}_o$
 to remain negative. In fact, conditions (\ref{dependence}) and 
(\ref{obtained}) are not compatible in general: their right-hand sides define 
two functions, $y(\omega_o)$ and $w(\omega_o)$, with $w$ more strongly 
dependent on $\omega_o$ than $y$. It drops to zero more quickly for large 
$\omega_o$ and diverges more quickly as $\omega_o$ becomes small. \\
Then, suppose to have a system at a sufficiently short orbital period and high 
$\alpha >1$ that spin-orbit coupling drives the two component stars apart. As 
the orbit widens, $(\alpha^2-1) \propto \omega^{-(2-\delta)}_o$, while the 
condition for orbital spin-down to continue requires a stronger, 
\textit{negative}, dependance. A point is reached, at which the degree of 
asynchronism is not sufficient anymore to sustain spin-orbit coupling against 
GWs and the orbit must start shrinking. Therefore, binary systems can only 
spin down via this mechanism for a finite time, after which they must shrink 
and eventually reach a state with $\alpha^{en}_{\infty} < 1$.

\end{document}